\documentclass[iop]{emulateapj}

\journalinfo{To Appear in ApJL}
\submitted{Received 2010 July 22; accepted 2010 October 12}

\shorttitle{Very-Hot Super-Earths}
\shortauthors{Schlaufman et al.}

\begin{document}

\title{A Population of Very-Hot Super-Earths in Multiple-Planet Systems Should
be Uncovered by {\it Kepler}}

\author{Kevin C. Schlaufman\altaffilmark{1,3}, D. N. C. Lin\altaffilmark{1,4},
and S. Ida\altaffilmark{2}}
\altaffiltext{1}{Astronomy and Astrophysics Department, University of
California, Santa Cruz, CA 95064, USA; kcs@ucolick.org and lin@ucolick.org}
\altaffiltext{2}{Tokyo Institute of Technology, Ookayama, Meguro-ku, Tokyo
152-8551, Japan; ida@geo.titech.ac.jp}
\altaffiltext{3}{NSF Graduate Research Fellow}
\altaffiltext{4}{Kavli Institute of Astronomy and Astrophysics, Peking
University, Beijing, China}

\begin{abstract}

\noindent
We simulate a {\it Kepler}-like observation of a theoretical exoplanet
population and we show that the observed orbital period distribution
of the {\it Kepler} giant planet candidates is best matched by an
average stellar specific dissipation function $Q_{\ast}'$ in the
interval $10^6 \lesssim Q_{\ast}' \lesssim 10^7$.  In that situation,
the few super-Earths that are driven to orbital periods $P < 1$ day by
dynamical interactions in multiple-planet systems will survive tidal
disruption for a significant fraction of the main-sequence lifetimes of
their stellar hosts.  Consequently, though these very-hot super-Earths
are not characteristic of the overall super-Earth population, their
substantial transit probability implies that they should be significant
contributors to the full super-Earth population uncovered by {\it Kepler}.
As a result, the CoRoT-7 system may be the first representative of a
population of very-hot super-Earths that we suggest should be found in
multiple-planet systems preferentially orbiting the least-dissipative
stellar hosts in the {\it Kepler} sample.

\end{abstract}

\keywords{planetary systems --- planets and satellites: formation ---
planets and satellites: individual: CoRoT-7 --- planet-star interactions}

\defcitealias{ida10}{IL10}

\section{Introduction}

NASA's {\it Kepler} mission
\citep{koc10,jen10a,cal10,bry10,bat10a,bat10b,haa10,jen10b} is
currently searching for transiting planets in photometric observations
of $\sim 156,000$ solar-type stars in a 115 deg$^2$ field of view in the
constellation Cygnus toward the Orion arm of our Galaxy.  {\it Kepler}
photometric observations will achieve 80 parts per million precision
over 80\%-90\% of the telescope field of view, sufficient to identify
the transit signal from an Earth-Sun equivalent system.  By the end of
its four-year mission, the combination of {\it Kepler}'s unprecedented
precision, sample size, and homogeneous selection will produce a near
ideal set of exoplanet detections to compare with theoretical models
of planet formation.  Though the preliminary exoplanet candidate list
announced in \citet{bor10} is incomplete, it is still a valuable constrain
on theoretical models of close-in planet formation.

In the core accretion model of close-in planet formation
\citep[e.g.,][]{pol96,ida04}, the cores of giant planets form near
the ice line of their parent protoplanetary disk.  The cores grow to
an isolation mass and accrete gas until they are massive enough to
open-up a gap in their parent disk.  The newly-formed giant planets
then Type II migrate into the close proximity of their host star and
stop inside the magnetospheric truncation radius of their parent
protoplanetary disk \citep[e.g.,][]{lin96}.  Lower-mass planets
in the close proximity of their host stars result from the inward
Type I migration \citep{war97} of planetary embryos that did not
reach the mass necessary to initiate runaway accretion of gas and
become giant planets.  These lower-mass close-in planets may be
hot Neptunes or icy super-Earths that formed near the ice line of
their parent protoplanetary disks and then migrated to just outside
the disk magnetospheric truncation radius \citep[e.g.,][]{mas06}.
Alternatively, they may be rocky Earth and super-Earth mass planets that
formed interior to the ice line and then migrated either singly or in
resonant planet convoys into the close proximity of their host star,
eventually stopping just outside the disk magnetospheric truncation
radius \citep[e.g.,][]{ida10,ogi10}.  Theoretical exoplanet population
synthesis (EPS) models \citep{ida08,ida10,mor09a,mor09b} in the core
accretion and migration paradigm have reproduced many trends in the
observed distribution of exoplanet properties, and comparisons between
observations and models are especially useful to constrain uncertain
model parameters \citep[e.g.,][]{sch09}.

Tidal interactions between close-in planets and their stellar
hosts contribute significantly to the dynamical evolution
of close-in planet systems after the dissipation of their
parent protoplanetary disks \citep[e.g.,][]{ras96,dob04}.
Tidal effects tend to circularize and shrink orbits to the
point that close-in planets suffer substantial mass loss and
eventually tidal disruption \citep[e.g.,][]{gu03,jac08a,jac09,li10}.
The efficiency and timescale for these processes are both uncertain
\citep[e.g.,][]{lev09} and highly dependent on the internal structure
of the close-in planet \citep[e.g.,][]{ogi04} and its stellar host
\citep[e.g.,][]{ogi07,bar09,bar10}.  Modeling of the properties of the
close-in planet candidates identified by {\it Kepler} will help resolve
some of these uncertainties.

In this letter, we combine an extended version of the EPS models of
\citet{ida10}--\citetalias{ida10} hereafter--with the mass--radius
relations of \citet{fort07} to extract the expected period--radius
distribution of the extended EPS.  We use Monte Carlo simulations in
concert with a detailed model of {\it Kepler} sensitivity and a simple
model of tidal evolution to determine the exoplanet period--radius
distribution that {\it Kepler} would observe in the extended EPS as a
function of the strength of tidal evolution.  We then compare the result
with the observed period distribution of {\it Kepler} transiting giant
planet candidates announced in \citet{bor10} to constrain the strength of
tidal evolution in the {\it Kepler} candidate planet sample.  We describe
our Monte Carlo simulations and {\it Kepler}-like observations of the
extended \citetalias{ida10} EPS in Section 2.  We discuss the results
and implications of our simulations in Section 3, and we summarize our
findings in Section 4.

\section{Analysis}

Transit observations directly measure exoplanet radii and orbital periods,
while radial velocity observations are necessary to confirm individual
transit detections and to determine exoplanet masses.  Unfortunately,
most exoplanet candidates identified by {\it Kepler} will orbit host stars
with faint apparent magnitudes.  As a result, radial velocity follow-up
will be telescope-time intensive for the more massive {\it Kepler}
exoplanet candidates and possibly too imprecise at faint magnitudes to
confirm lower-mass candidates.  Consequently, it becomes important to
carefully consider what insight can be gained from the exoplanet period
distribution derived purely from {\it Kepler} transit photometry.

To that end, we use a Monte Carlo simulation to systematically observe
an extended version of the \citetalias{ida10} EPS to as much as possible
match {\it Kepler} observations of the Milky Way's exoplanet population.
In this case, we have extended the \citetalias{ida10} EPS such that
the distribution of host stellar mass in the 10$^3$ systems in the EPS
matches the distribution of host stellar mass expected in {\it Kepler}
observations.  That is, we match the distribution of host stellar mass in
the extended \citetalias{ida10} EPS to the mass distribution expected in
a {\it Kepler}-like observation of the \citet{rei05} solar neighborhood
luminosity function, given {\it Kepler}'s sensitivity as a function of
host spectral type and apparent magnitude.

The first step in our calculation is to approximate the radius of each
planet in the extended \citetalias{ida10} EPS.  The \citetalias{ida10}
models include the asymptotic semimajor axis, rock mass, ice mass, and
gas mass of every planet.  From those quantities, we use the mass--radius
relations given in \citet{fort07} to assign radii to giant planets as a
function of semimajor axis, solid mass, and gas mass and to terrestrial
planets as a function of solid mass and composition.

We then do 100 iterations of a Monte Carlo simulation in which we assign
each planetary system in the extended IL10 EPS a random host stellar mass,
age, and apparent magnitude, as well as a random orbital inclination.
We assign each eccentric planet a random argument of periastron.
We first randomly assign spectral types and apparent magnitudes to
the host stars of the extended exoplanet population.  We use the solar
neighborhood luminosity function given in Table 8.3 of \citet{rei05}
to determine the number density of FGK stars as a function of spectral
type and apparent magnitude; we use this information to determine the
probability that a randomly selected star from the {\it Kepler} field
is of a given spectral type and apparent magnitude.  The spectral type
of the assigned host star determines the host stellar mass and radius.

To account for the effect of tidal evolution on the observability of
the extended \citetalias{ida10} exoplanet population, we exploit the
fact that the two-Gyr moving-average smoothed star formation history
in the Milky Way has been more or less constant over the last $\sim10$
Gyr \citep[e.g.,][]{roc00a,roc00b}.  As a result, the unknown age of
a randomly selected star from a magnitude-limited survey of Milky Way
disk stars with main-sequence lifetime $\tau_{\ast}$ short relative to
that $\sim10$ Gyr interval should be distributed more or less uniformly
between zero and its main-sequence lifetime.  Magnitude-limited transit
surveys like {\it Kepler} are biased toward stars with $M_{\ast}
\gtrsim 1~M_{\odot}$, and those stars have main-sequence lifetimes
$\tau_{\ast} \lesssim 10$ Gyr.  Consequently, the unknown system
age $\tau_{\mathrm{sys}}$ of a typical candidate exoplanet system
identified by {\it Kepler} in transit with host stellar mass $M_{\ast}$
should be well-approximated as uniformly distributed in the interval
$0 \leq \tau_{\mathrm{sys}} \leq \tau_{\ast}(M_{\ast})$.  For that
reason, in the Monte Carlo we randomly sample the age of each exoplanet
system from a uniform distribution between zero and the main-sequence
lifetime of its stellar host.  For Sun-like stars $L_{\ast}(M_{\ast})
= L_{\odot}~(M_{\ast}/M_{\odot})^{3.5}$ \citep[e.g.,][]{pop80}, and
the total amount of hydrogen available for fusion is proportional
to $M_{\ast}$.  The main-sequence lifetime is roughly then
$\tau_{\ast}(M_{\ast}) = \tau_{\odot}~M_{\ast}/L_{\ast}(M_{\ast}) =
10~(M_{\ast}/M_{\odot})^{-2.5}$ Gyr.  We consider an exoplanet observable
if its randomly determined age $\tau_{\mathrm{sys}}$ is less than
the timescale for the tidal evolution of its orbit to move it within
$1~R_{\ast}$ of its stellar host $\tau_{\mathrm{dis}}$ at which point
we assume that the planet is tidally disrupted and no longer observable
\citep[e.g.,][]{san98}.  In other words, a planet is only observable
if $\tau_{\mathrm{sys}} < \tau_{\mathrm{dis}}$.  We define the time to
disruption as \citep[e.g.,][]{ibg09}

\begin{eqnarray}\label{eq1}
\tau_{\mathrm{dis}} & = & \frac{4}{117} \frac{a_{0}^{13/2}}{G^{1/2}}
\frac{M_{\ast}^{1/2}}{M_{p}} \frac{Q_{\ast}'}{R_{\ast}^5}
\left[1-(R_{\ast}/a_0)^{13/2}\right] \mathrm{,}
\end{eqnarray}

\noindent
where $a_0$ is the initial semimajor axis of the planet before tidal
evolution, $Q_{\ast}'$ is the specific dissipation function of the host
star, $G$ is Newton's gravitational constant, $M_p$ is the mass of the
planet, and $R_{\ast}$ is the radius of the host star. The expression
given in \citet{ibg09} is strictly valid only for circular orbits, so
we assume that the timescale for eccentricity damping is short relative
to $\tau_{\mathrm{dis}}$.

As a result of the assumptions we make in our treatment of
tidal evolution, we assume that the all planets from the extended
\citetalias{ida10} population with orbital period $P < 10$ days are on
circular orbits.  We use the eccentricity distribution generated from the
EPS models for longer-period systems.  We sample the orbital inclination
and argument of periastron from the standard random distributions of
those quantities.  We adopt the analytic formulae given in \citet{sea03}
and \citet{ford08} to determine the transit depth and transit duration
of every planet found to transit.  Finally, we use a detailed model of
{\it Kepler} sensitivity (D. Koch et al., private communication) that
gives the threshold detectable exoplanet radius as a function of host
spectral type, host apparent magnitude, transit depth, transit duration,
and number of transits over a given duration of observation to determine
which of the simulated transiting exoplanets would be detectable in the
first 43 days of {\it Kepler} data.

\begin{figure}
\plottwo{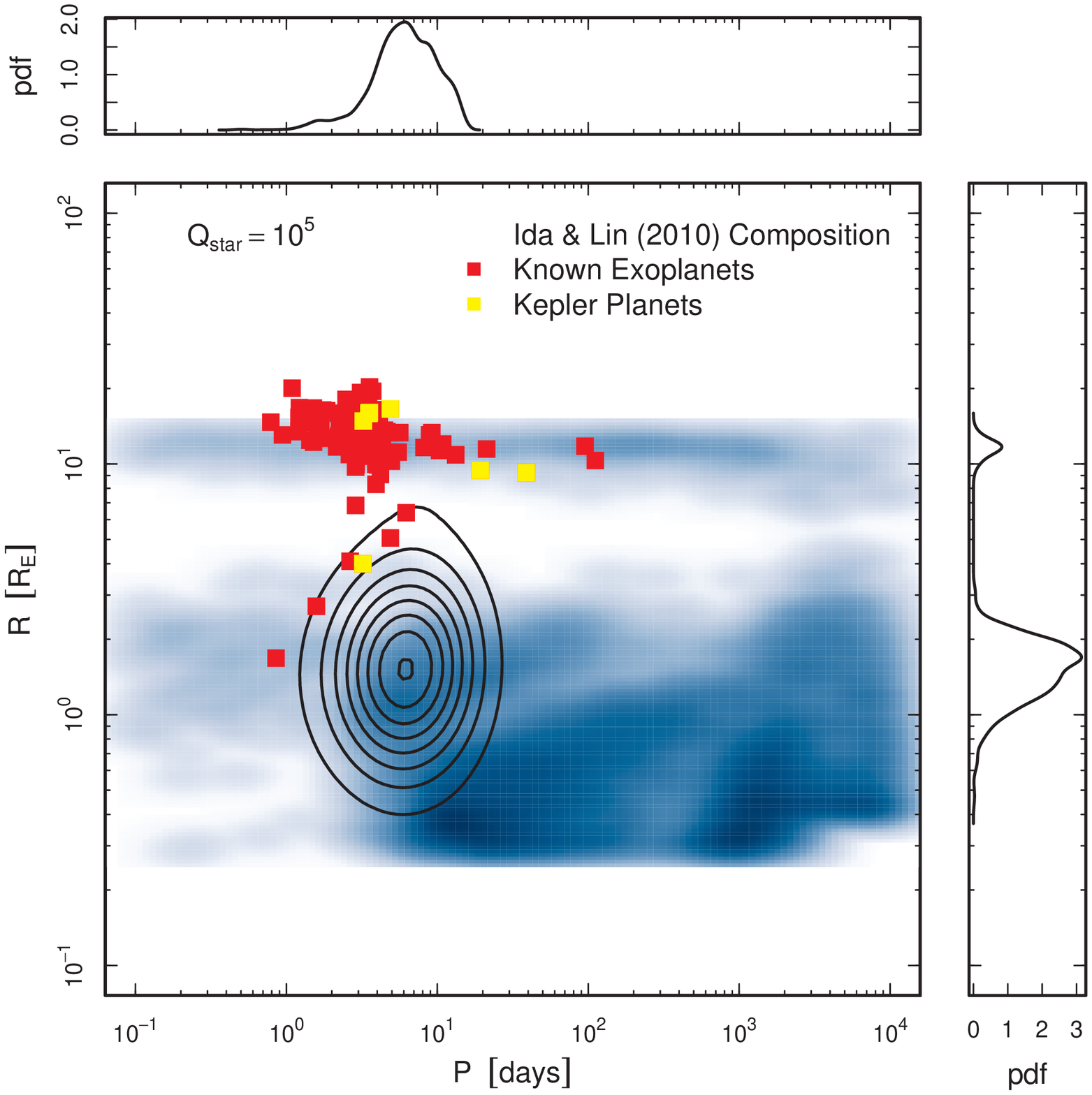}{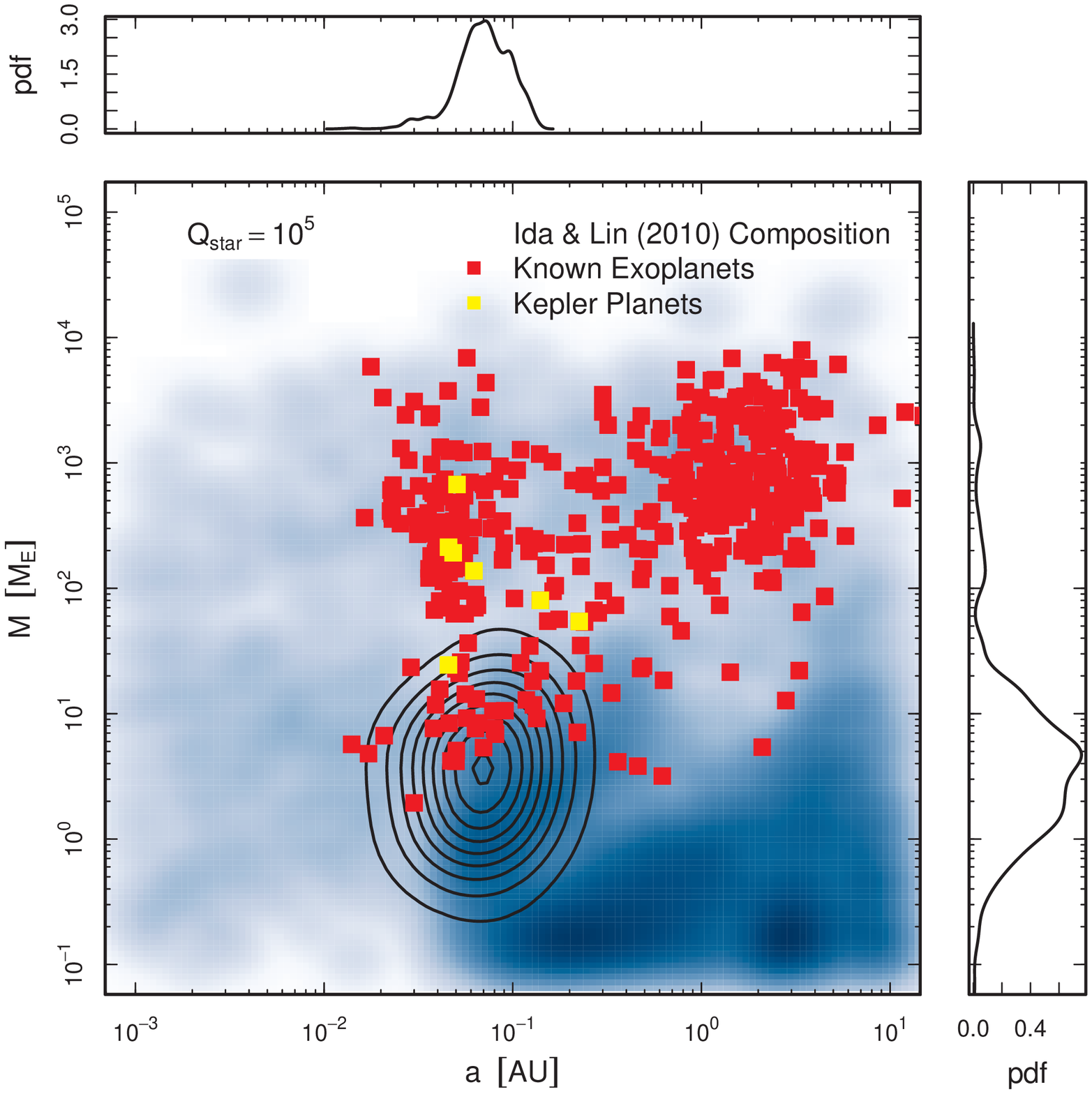}
\caption{Simulated {\it Kepler} observation of the extended
\citetalias{ida10} EPS models assuming the mass--radius relation from
\citet{fort07} and strong tidal evolution (i.e., $Q_{\ast}' = 10^5$).
The blue shading shows the underlying initial extended EPS before tidal
evolution, while the black contours show the distribution of exoplanet
properties including tidal evolution that would have been observable by
{\it Kepler} in its first 43 days of science operation.  We plot known
exoplanets as red squares and confirmed {\it Kepler} planets as yellow
squares.  \emph{Left}: Period--radius plane with marginal distributions.
In this case, the orbits of giant planets quickly degrade due to strong
tidal evolution.  \emph{Right}: Semimajor axis--mass plane with marginal
distributions.\label{fig01}}
\end{figure}

\begin{figure}
\plottwo{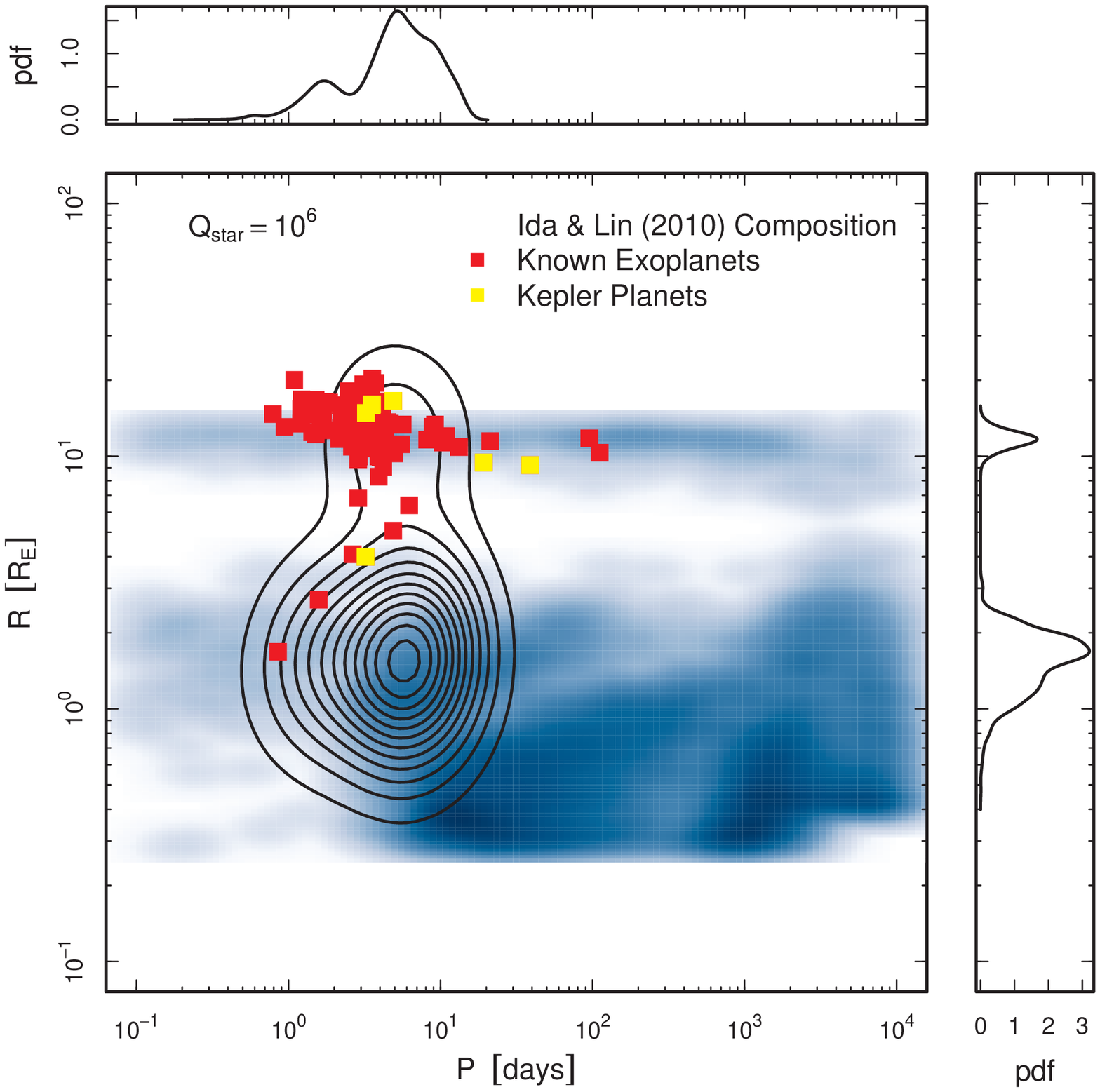}{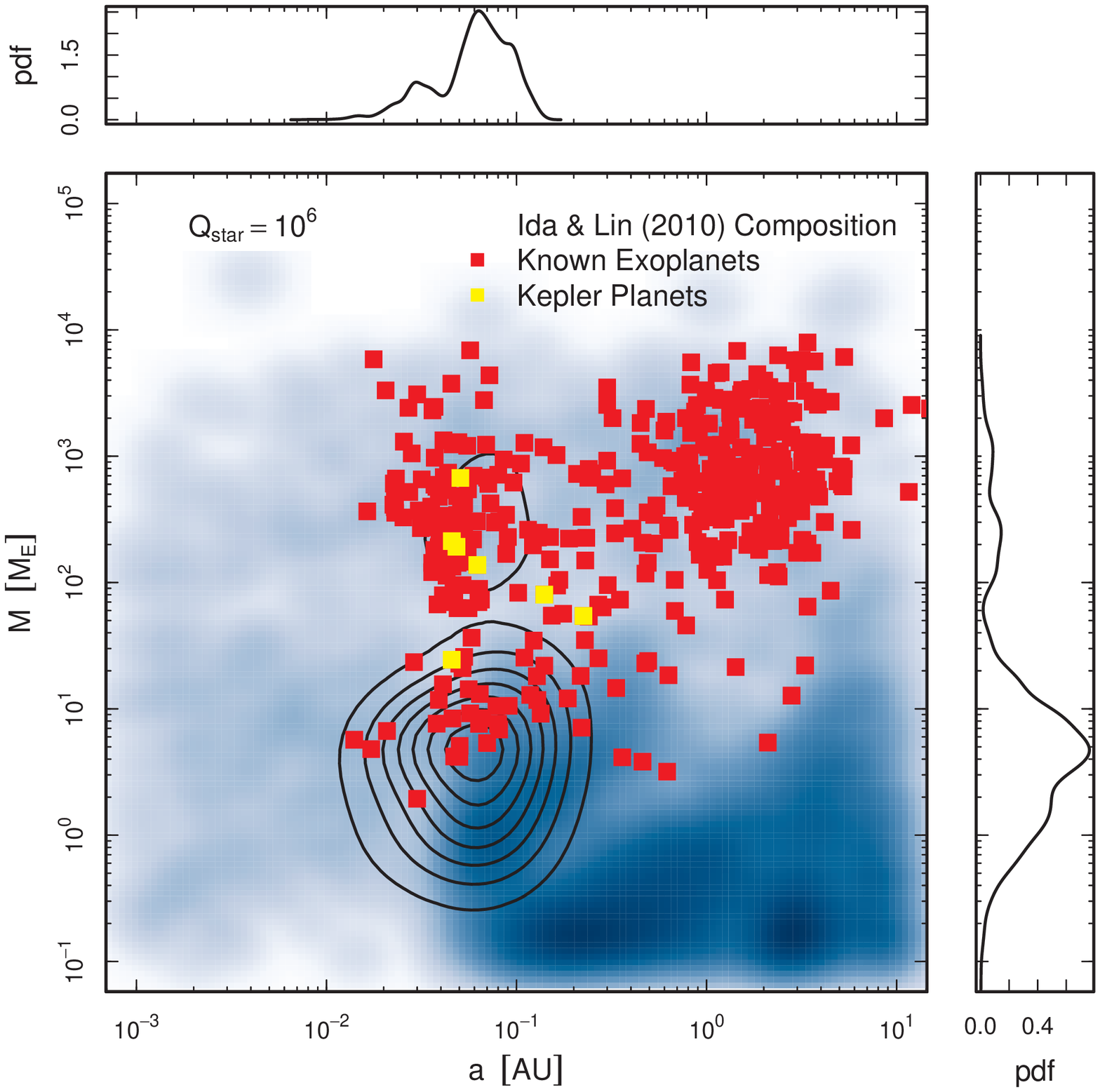}
\caption{Same as Figure~\ref{fig01} but assuming moderate tidal evolution
(i.e., $Q_{\ast}' = 10^6$).  In this case, the moderate tidal evolution
allows some giant planets to survive.\label{fig02}}
\end{figure}

\begin{figure}
\plottwo{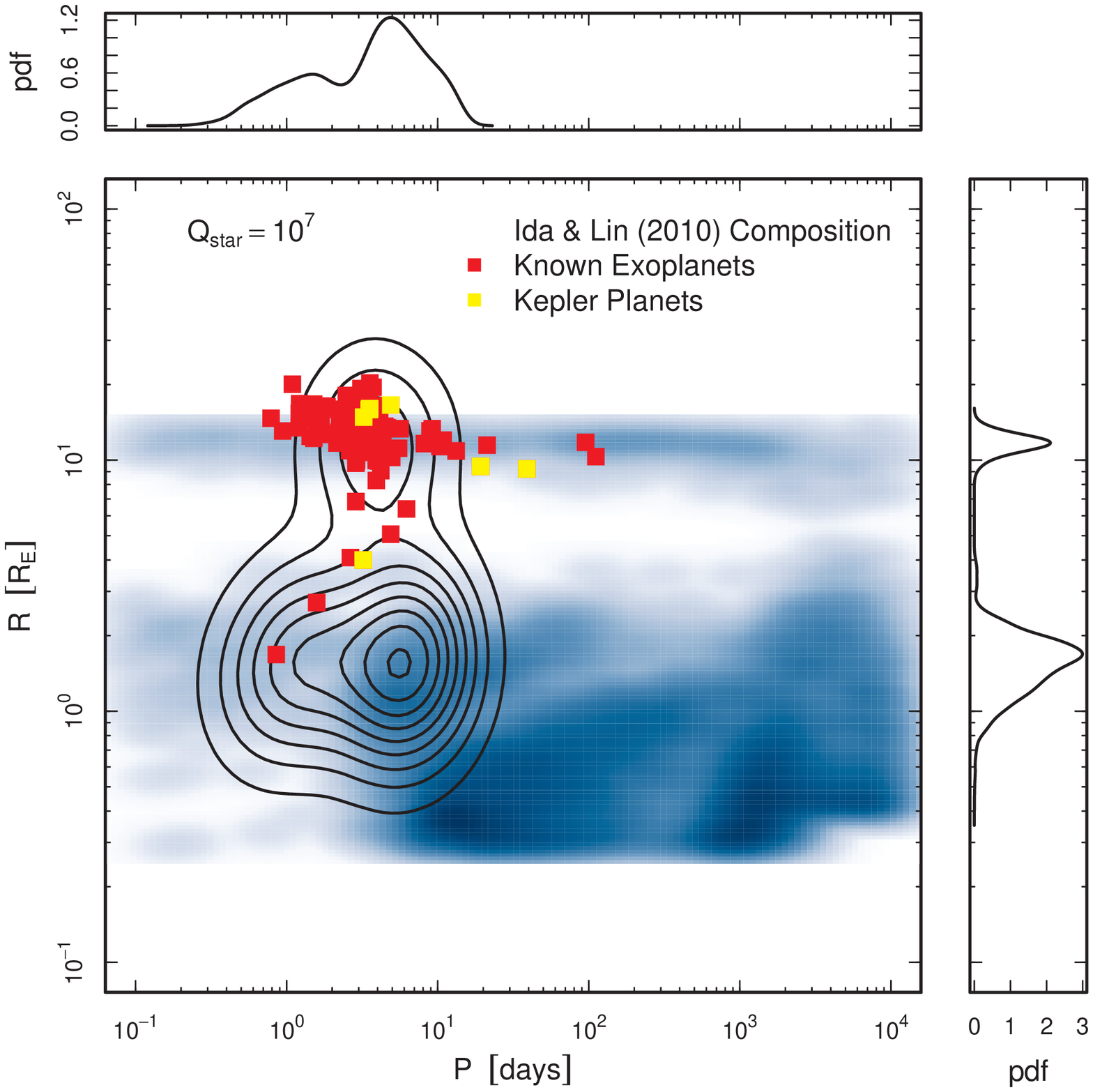}{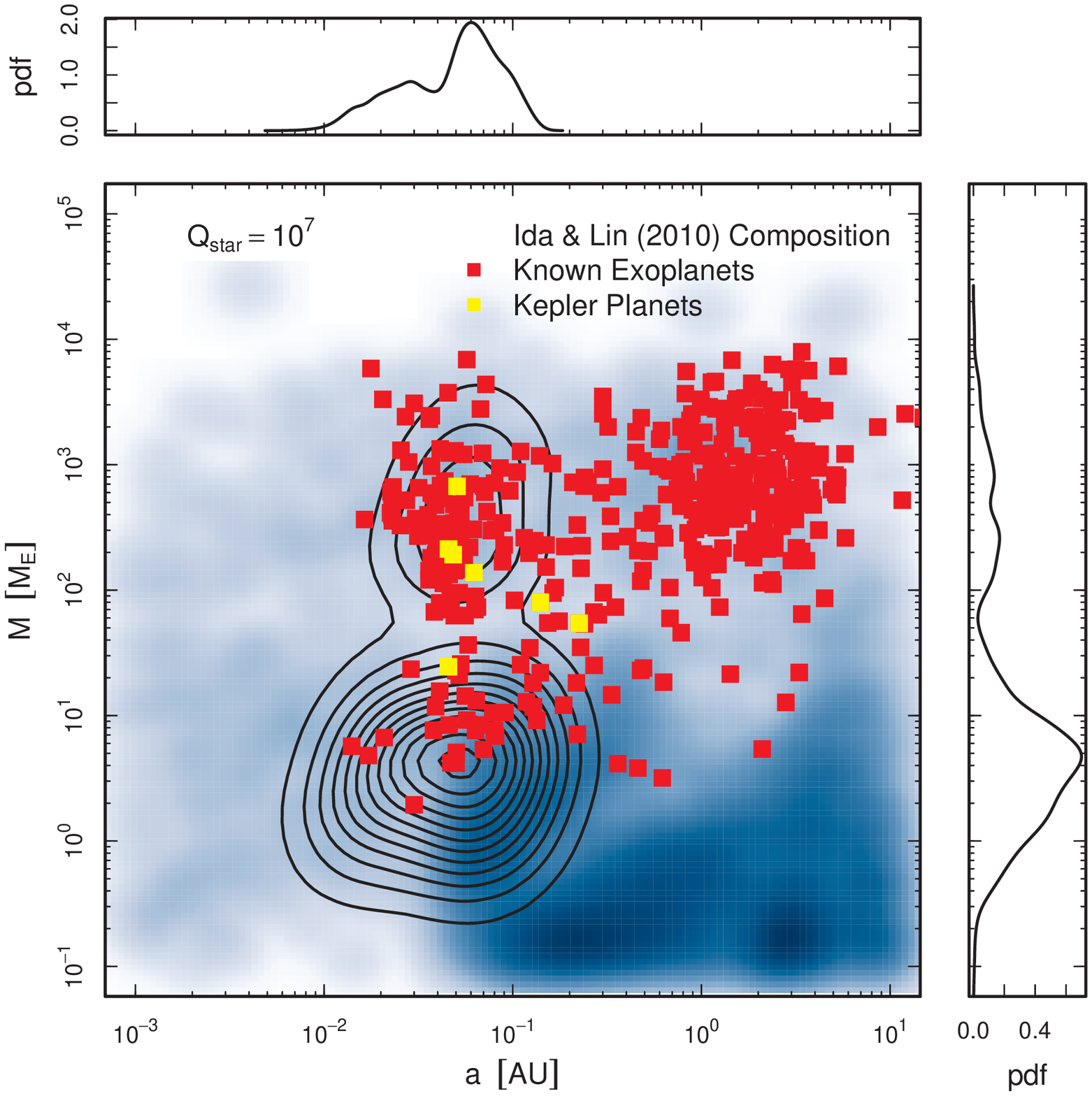}
\caption{Same as Figure~\ref{fig01} but assuming weak tidal evolution
(i.e., $Q_{\ast}' = 10^7$).  In this case, the weak tidal evolution
allows many giant planets to survive.  Interestingly, the observed period
distribution of super-Earths extends to $P < 1$ day, as super-Earths
scattered to short-period orbits through dynamical interactions in
multiple-planet systems would persist for a significant fraction of the
main-sequence lifetime of their host stars.  These very-hot super-Earths
should occur in multiple-planet systems, and should preferentially
orbit the least-dissipative host stars in the {\it Kepler} sample (i.e.,
those stars with $M_{\ast} \gtrsim 1.25~M_{\odot}$, convective cores,
and radiative envelopes).\label{fig03}}
\end{figure}

\begin{figure}
\plottwo{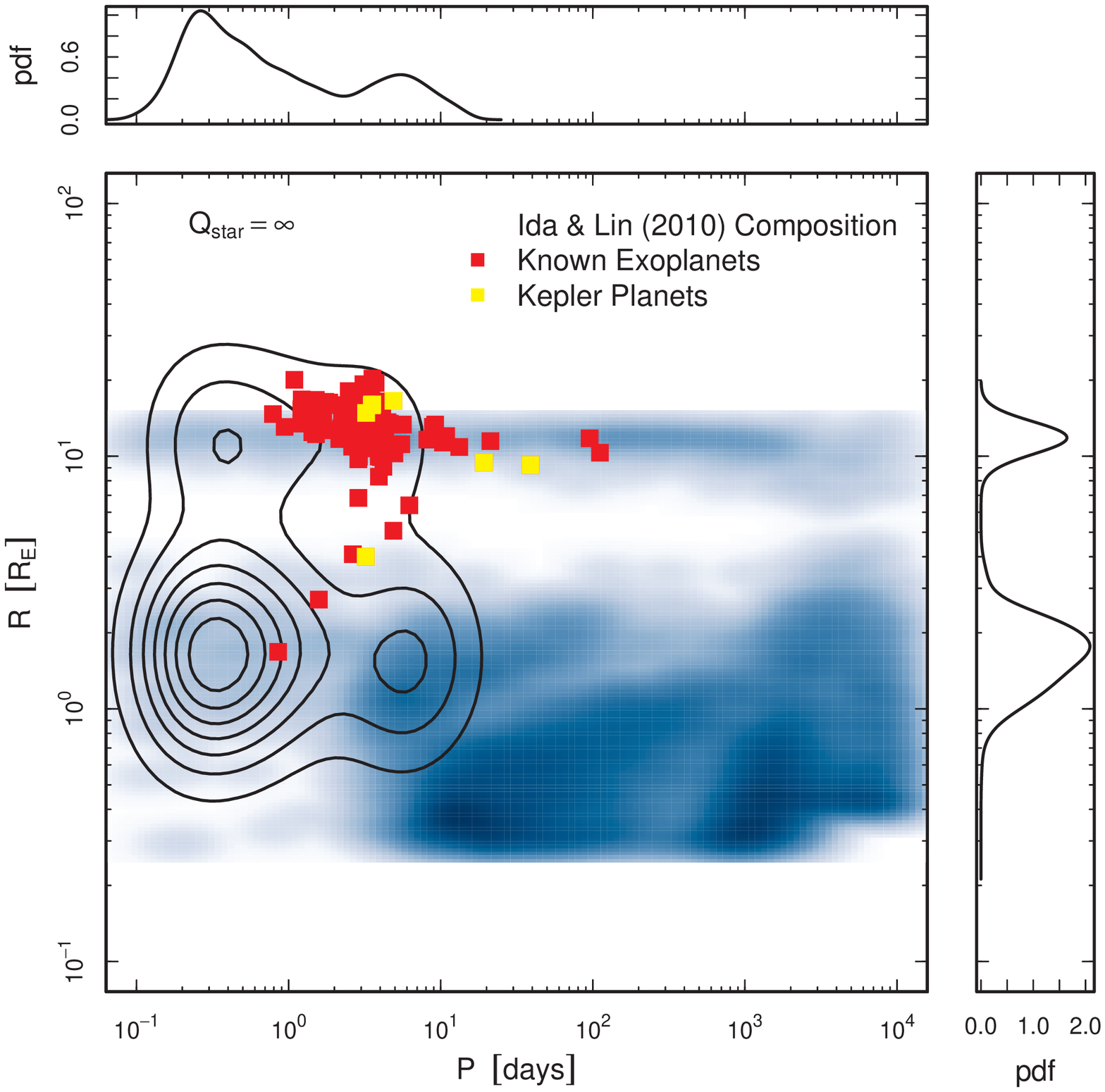}{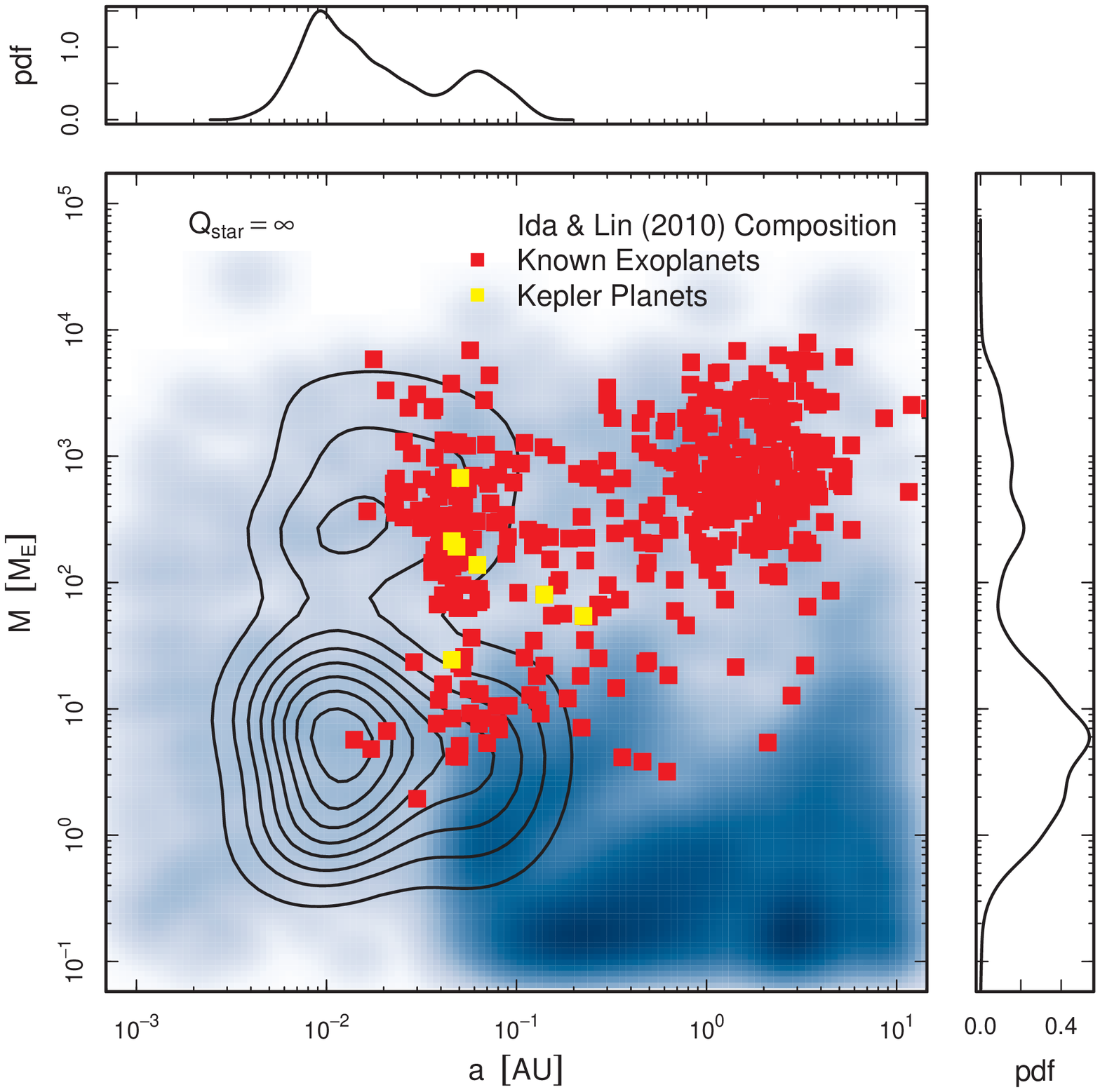}
\caption{Same as Figure~\ref{fig01} but assuming no tidal evolution
(i.e., $Q_{\ast}' = \infty$).  In this case, the lack of tidal evolution
allows all very-short period systems produced by dynamical interactions
to survive.\label{fig04}}
\end{figure}

We save the results of this iteration, and repeat our Monte Carlo
simulation 100 times to average over all of the given distributions
to generate a prediction for the characteristics of the planetary
systems {\it Kepler} would likely identify in an exoplanet population
matching the extended \citetalias{ida10} EPS.  We carry out this
Monte Carlo simulation for four fiducial values of $Q_{\ast}'$: 10$^5$
(strong tidal evolution), 10$^6$ (moderate tidal evolution), 10$^7$
(weak tidal evolution), and $\infty$ (no tidal evolution).  We present
the results of our simulations for each assumed value of $Q_{\ast}'$
in Figures~\ref{fig01} to \ref{fig04}.

\section{Discussion}

The exoplanet period--radius distribution that results from a {\it
Kepler}-like observation of the extended \citetalias{ida10} EPS is a
strong function of $Q_{\ast}'$.  In the most dissipative case $Q_{\ast}'
= 10^5$, hot Jupiters are quickly disrupted and the period distribution
of giant planets is flat with a peak at $P \approx 10$ days.  At the
same time, the handful of planets at very short period $P < 1$ day
produced by dynamical interactions between planets in multiple-planet
systems after the dissipation of their parent protoplanetary disk are
tidally disrupted and unobservable in a {\it Kepler}-like observation.
In the less dissipative cases $Q_{\ast}' = 10^6$ and $Q_{\ast}' = 10^7$,
hot Jupiters survive for a significant fraction of the main-sequence
lifetime of their host star and the period distribution of giant planets
peaks in the range $3 < P < 5$ days in agreement with the giant planet
period distribution in \citet{bor10}.  In those cases, the observed
period distribution of super-Earths would have a long tail toward periods
shorter than one day.  In the case with no tidal dissipation $Q_{\ast}'
= \infty$, hot Jupiters survive for the entire main-sequence lifetime
of their stellar hosts and the period distribution of giant planets
is a monotonically decreasing function of period.  As a result, a {\it
Kepler}-like observation would be dominated by planets with periods less
than one day.  Unlike the \citetalias{ida10} models, our calculations
suggest that in the case of weak or no tidal evolution super-Earths
should be observed to have a shorter average period than giant planets.

\begin{figure}
\epsscale{.80}
\plotone{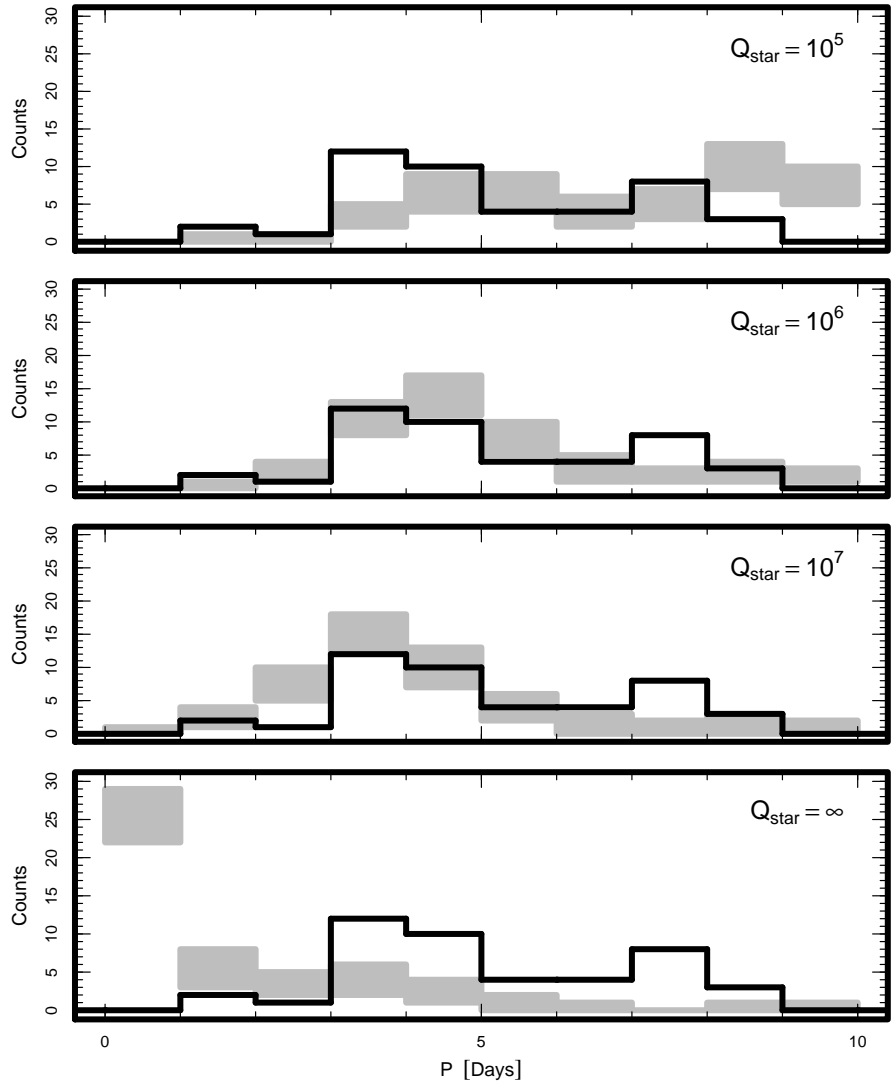}
\caption{Histograms describing the period distribution of short-period
giant planets.  The black curve represents the period distribution of
short-period giant planets (e.g., planets with $R_{p} > 0.5~R_J$ and $P <
10$ days) from \citet{bor10}.  The gray rectangles represent the one-sigma
range in the period distribution of short-period giant planets (e.g.,
planets with $M_{p} > 50~M_{\oplus}$ and $P < 10$ days) expected from
a {\it Kepler}-like observation of the extended \citetalias{ida10} EPS.
Each panel corresponds to different assumptions for the strength of tidal
evolution in the simulated {\it Kepler} observation of the extended
\citetalias{ida10} EPS.  Given the extended EPS, the first 43 days of
{\it Kepler} science data already rule-out both strong tidal evolution
and no tidal evolution; the current data favors $10^6 \lesssim Q_{\ast}'
\lesssim 10^7$.\label{fig05}}
\end{figure}

In Figure~\ref{fig05} we compare the giant planet period distribution
that results from our simulated {\it Kepler}-like observation of the
extended \citetalias{ida10} EPS with the {\it Kepler} planet-candidate
period distribution from \citet{bor10}.  Massive close-in planets are most
sensitive to tidal evolution, so we only compare the period distribution
of giant planets from the extended \citetalias{ida10} EPS more massive
than $50~M_{\oplus}$ with those {\it Kepler} planet candidates with radii
greater than 0.5 Jupiter radii.  The {\it Kepler} planet candidates from
\citet{bor10} are not necessarily representative of the properties of
the still-embargoed full sample of {\it Kepler} planet candidates.
At the same time, the announced candidates are very likely those
candidates unsuitable for radial velocity follow-up.  For the massive
planet candidates at least, that only indicates they orbit apparently
faint host stars.  Properties like the average metallicity of a stellar
population are not a strong function of Galactocentric radius, so there is
no reason to believe that the subsample of planets that orbit apparently
faint stars at distances $d \sim 500$ pc from the Sun is systematically
different than the subsample of planets that orbit apparently bright stars
at distances $d \sim 200$ pc from the Sun.  For those reasons, we argue
that a comparison of the period distribution of the announced {\it Kepler}
exoplanet candidates with the period distribution expected from a {\it
Kepler}-like observation of the extended \citetalias{ida10} EPS under
different assumptions for the strength of tidal evolution is meaningful.

The period distribution of the {\it Kepler} planet candidates is
inconsistent with both strong tidal evolution and no tidal evolution
given the extended \citetalias{ida10} EPS.  The observed distribution
is consistent with the two intermediate values of $Q_{\ast}' = 10^6$ and
$Q_{\ast}' = 10^7$, with a better match provided by $Q_{\ast}' = 10^7$.
This measurement is also consistent with the typically quoted $Q_{\ast}'
= 10^6$ from the literature \citep[e.g.,][]{ogi07}.

The average specific dissipation function $Q_{\ast}' = 10^7$ suggested
by our analysis implies that though very-hot super-Earths are not
common, their substantial transit probability ensures that they will
be readily observable by {\it Kepler}.  We expect that about 10\% of
detected planets with $P < 10$ days and $M_{p} < 10~M_{\oplus}$ should
be very-hot super-Earths with $P < 1$ day.  Very-hot super-Earths are
produced by dynamical interactions in multiple-planet systems, and we
suggest that very-hot super-Earths systems identified in transit surveys
should frequently have observable companions \citep[e.g.,][]{mar04}.
Indeed, the very-hot super-Earth CoRoT-7b \citep{leg09} is in a confirmed
multiple system \citep{que09} with the recently suggested possibility
of a third planet \citep{hat10}.  Very-hot super-Earths should also
preferentially be found around Sun-like stars with $M_{\ast} \gtrsim
1.25~M_{\odot}$, as these stars are likely less dissipative than stars
with $M_{\ast} \approx 1~M_{\odot}$ \citep{bar09}.  At the same time,
there are other explanations for the properties of the CoRoT-7 system
that match the observations \citep[e.g.,][]{jac10}.

There are many limitations to our approach.  We use only a single value
of $Q_{\ast}'$ to model the strength of tidal evolution regardless
of host star age or spectral type, when is there is evidence from
both observation \citep[e.g.,][]{sch10,win10} and theoretical models
\citep[e.g.,][]{bar09,bar10} that $Q_{\ast}'$ is a function of host
star age and stellar structure, as well as exoplanet orbital period.
This issue will be resolved in the future when the full list of {\it
Kepler} planet candidates is announced, as the increased host star
statistics will allow us to do similar calculations in bins of host star
effective temperature and thereby constrain $Q_{\ast}'$ as a function
of $T_{\mathrm{eff}}$ and therefore stellar structure.  We do not
simultaneously evolve the orbit and radius of the exoplanet population
\citep[e.g.,][]{jac08b,ibg09,mil09}, and the migration stopping conditions
discussed in Section 2.5 of \citetalias{ida10} are also uncertain.

In \citetalias{ida10}, planets that were massive enough to open-up
a gap in their parent protoplanetary disk were stopped just inside
the disk magnetospheric truncation radius (taken to be at $P = 2$
days), while planets that were not massive enough to open-up a gap
were stopped just outside of the magnetospheric truncation radius
(taken to be at $P = 3$ days).  Though these stopping conditions are
dependent on the uncertain properties of the parent protoplanetary
disk \citep[e.g.,][]{lin96,mas06,ogi10}, they are self-consistent in
that they describe where an exoplanet of a given mass will stop if the
stellar magnetic field and disk mass-accretion rate have a given value.

\section{Conclusion}

We coupled a detailed model of {\it Kepler} sensitivity with the results
of the extended \citetalias{ida10} EPS, a simple model of tidal evolution,
and published mass--radius relations for both terrestrial and giant
planets to determine the period--radius distribution expected under the
extended \citetalias{ida10} EPS.  We found that the period distribution
of close-in giant planets is a strong function of the strength of
tidal evolution parametrized as $Q_{\ast}'$, and we showed that the
period distribution of the {\it Kepler} planet candidates announced
by \citet{bor10} is best matched by the extended \citetalias{ida10}
EPS with $10^6 \lesssim Q_{\ast}' \lesssim 10^7$.  In that case, there
exists a population of very-hot super-Earths with periods less than one
day that results from dynamical interactions in multiple-planet systems.
Though these very-hot super-Earths are rare, their relatively high
transit probability and long time-scale for tidal disruption when $10^6
\lesssim Q_{\ast}' \lesssim 10^7$ indicate that they will be observable
by {\it Kepler}.  The predicted very-hot super-Earths are analogous to
the CoRoT-7 system, and we suggest that {\it Kepler} should find many
such very-hot super-Earths in multiple-planet systems preferentially
around stellar hosts with convective cores and radiative envelopes.

\acknowledgments
We thank Jonathan Fortney and Neil Miller for useful conversation and the
anonymous referee for helpful suggestions.  We are especially grateful
to David Koch and the {\it Kepler} team for providing us with a detailed
model of {\it Kepler} sensitivity.  This research has made use of NASA's
Astrophysics Data System Bibliographic Services.  This material is based
upon work supported under a National Science Foundation Graduate Research
Fellowship, NASA (NNX07A-L13G, NNX07AI88G, NNX08AL41G, NNX08AM84G), NSF
(AST-0908807), and JSPS.

{\it Facilities:} \facility{Kepler}

\clearpage

\clearpage

\clearpage

\clearpage

\clearpage

\end{document}